\documentclass[preprint,journal]{vgtc}            

\onlineid{0}

\vgtccategory{Research}

\vgtcpapertype{please specify}

\title{Authoring Narrative Visualization in Motion:\\Visual Storytelling in Swimming Videos}

\author{%
  \authororcid{Junhao Zhao}{0000-0002-3271-6455},
  \authororcid{Romain Vuillemot}{0000-0003-1447-6926}, 
  \authororcid{Petra Isenberg}{0000-0002-2948-6417}, and 
  \authororcid{Lijie Yao}{0000-0002-4208-5140}
}

\authorfooter{
  \item
  	Junhao Zhao and Lijie Yao are with Xi'an Jiaotong-Liverpool University.
  	E-mail: nemozjh@hotmail.com, yaolijie0219@gmail.com.
  \item
  	Romain Vuillemot is with Universit{\'e} de Lyon, {\'E}cole Centrale de Lyon, CNRS, UMR5205, LIRIS, F-69134, France. E-mail: romain.vuillemot@ec-lyon.fr.
  \item
    Petra Isenberg is with Universit{\'e} Paris-Saclay, CNRS, Inria. E-mail: petra.isenberg@inria.fr.
  \item
    Corresponding author: Lijie Yao.
}

\abstract{
  We investigate how to support authoring narrative visualizations in motion in sports videos, drawing on automated data preparation, systematic analysis, technology probe design, and evaluation, using swimming races as a case study. 
  Sports videos are widely broadcast and shared across social media, where content creators increasingly seek to present and explain complex events to general audiences. Visualization in motion has been explored as an efficient way to embed data into videos and to move with the data referents, providing additional information and helping audiences understand races. However, existing approaches primarily focus on embedding visualizations in videos, lacking exploration of how to support authoring narratives that coordinate views, data, and temporal progression to explain the unfolding races.
  To address this gap, we use swimming videos as an ideal case for exploration, as swimming is a sport with rich, dynamic data and visualizations in practice. We develop an automated pipeline that extracts structured data from videos, derive narrative constructs through observational analysis of sports broadcasts, and design a technology probe that supports authoring using data prepared by our pipeline and narrative constructs derived from our observations. We evaluate our approach with experienced content creators and/or graphic designers to examine the benefits and challenges of authoring narrative visualizations in motion. 
  All supplemental materials are described in the Supplemental Material Pointers section and are on OSF: \href{https://osf.io/bq47n/}{\texttt{osf\discretionary{}{.}{.}io\discretionary{/}{}{/}bq47n\discretionary{/}{}{/}}}.
}


\teaser{
  \centering
  \includegraphics[width=\linewidth, alt={A view of a city with buildings peeking out of the clouds.}]{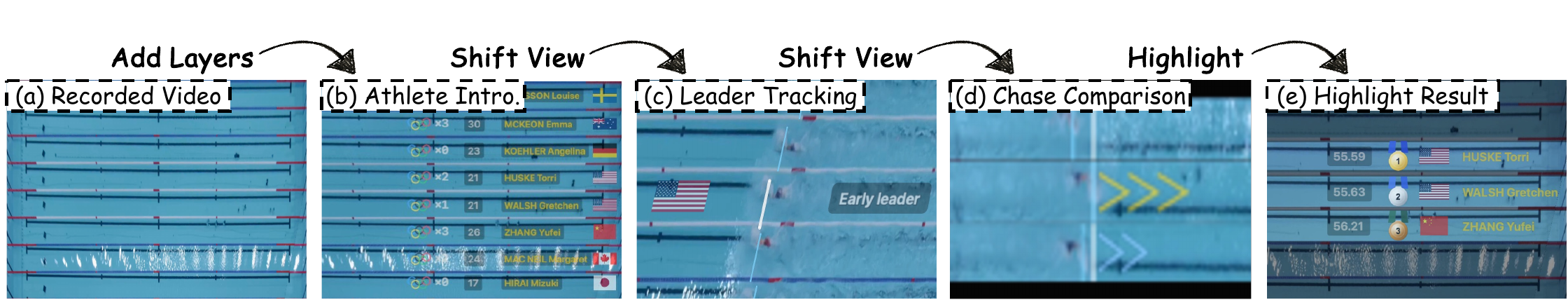}
  \caption{%
    Representative frames of narrative storytelling with visualizations in motion in a swimming race video authored with \SCe.  
    Starting from (a)~our swimming video recorded under authorization at 2024 Paris Olympics, the author adds data layers to create (b)~an \emph{Athlete Introduction} in \emph{Overview}, shifts to (c)~\emph{Tracking} to follow the early leader, then to (d)~\emph{Comparison} to reveal subtle differences in the chase, and finally highlights race results, including winners' completion times and information.
  }
  \label{fig:teaser}
}

\graphicspath{{Figures/}{figures/}{pictures/}{images/}{./}} 

\usepackage{tabu}                      
\usepackage{booktabs}                  
\usepackage{lipsum}                    
\usepackage{mwe}                       
\usepackage[dvipsnames, svgnames, table]{xcolor}
\usepackage{mathptmx}                  
\usepackage{eurosym}                   
\usepackage{wasysym}                   
\usepackage{adjustbox}                 
\usepackage{mathptmx}                  
\usepackage{tikz}
\usepackage{xurl}
\usepackage{xcolor}
\usepackage{ccicons}
\usepackage{cite}

\definecolor{myorange}{HTML}{F2AA84}
\definecolor{lightorange}{HTML}{F8D3C2}

\newcommand*{\orangecircle}[2][]%
  {\tikz[baseline=(C.base)]{%
    \node[inner sep=0pt] (C) {\vphantom{1g}};
    \node[draw, circle,fill=myorange,draw=myorange, inner sep=.15pt, yshift=.3pt] 
        at (C.center) {\vphantom{1g}};
    \node[inner sep=0pt] (C) {\vphantom{1g}#2};
        }
        }

\makeatletter
\DeclareRobustCommand\onedot{\futurelet\@let@token\@onedot}
\def\@onedot{\ifx\@let@token.\else.\null\fi\xspace}
\def\eg{\emph{e.g}\onedot} 
\def\ie{\emph{i.e}\onedot}

\def\etal{\emph{et al}\onedot}
\makeatother

\usepackage{siunitx}
\DeclareSIUnit\pixel{px}

\newcommand{\snd}[1]{\SI{#1}{\second}}
\newcommand{\mins}[1]{\SI{#1}{\minute}}
\newcommand{\m}[1]{\SI{#1}{\meter}}

\newcommand{\SCe}{\textit{SwimComposer}}

\usepackage{transparent}
\definecolor{darkBlue}{HTML}{3C6CA8}
\definecolor{middleBlue}{HTML}{8ECCEA}
\definecolor{lightBlue}{HTML}{DFEFF6}
\definecolor{beige}{HTML}{FFFCF6}
\definecolor{darkGrey}{HTML}{424242}
\definecolor{middleGrey}{HTML}{9D9D9D}
\definecolor{lightGrey}{HTML}{D3D3D3}
\definecolor{lightOrange}{HTML}{FCDE7F}
\definecolor{lightPurple}{HTML}{e6d1ff}

\newlength{\myLength}	
\newlength{\mytextsize}

\usepackage{tikz}
\newcommand{\percentagebar}[1]{\begin{tikzpicture}[baseline = .08\mytextsize]%
\draw[darkGrey, fill=white, thin,opacity=1] [yshift=0.7pt] (0,0) rectangle (\myLength,.6\mytextsize);
\draw[darkGrey, fill=darkGrey, thin,opacity=1] [yshift=0.7pt] (0,0) rectangle (#1 \myLength,.6\mytextsize);
\end{tikzpicture}
}
\usetikzlibrary{tikzmark}

\newcommand{\roundedDClabel}[1]{%
    \tikz[baseline=(char.base)] 
        \node[fill=darkBlue, 
              rounded corners=1.5pt, inner sep=1.5pt, 
              text=white, font=\sffamily\normalfont] (char) {#1};%
}

\DeclareRobustCommand{\panellabel}[1]{%
    \tikz[baseline=(char.base)]{
        \node[
            fill=lightorange,
            draw=lightorange,
            rounded corners=1.5pt,
            inner sep=1.5pt,
            text=black,
            font=\normalfont
        ] (char) {#1};
    }%
}

\DeclareRobustCommand{\dottedlabel}[1]{%
    \tikz[baseline=(char.base)]{
        \node[
            draw=black,
            dash pattern=on 1pt off 1pt,
            rounded corners=1.5pt,
            inner sep=1pt,
            text=black,
            font=\normalfont
        ] (char) {#1};
    }%
}

\DeclareRobustCommand{\solidlabel}[1]{%
    \tikz[baseline=(char.base)]{
        \node[
            draw=black,
            rounded corners=1.5pt,
            inner sep=1pt,
            text=black,
            font=\normalfont
        ] (char) {#1};
    }%
}

\DeclareRobustCommand{\plabel}[1]{%
    {\normalfont #1}%
}

\newcommand{\revision}[1]{{\color{black}#1}}

\begin{document}


\firstsection{Introduction}

\maketitle

Sports videos are widely consumed, edited, and shared across social media due to the public's interest in games and competitions. Content creators thus increasingly seek to present and explain sports events to general audiences, not only as raw footage but also with related data and visual effects that capture audiences' attention, support their following of an event, and deepen their understanding.
Recent work has explored embedding visualizations, visual analytics, and visual effects into videos across sports, including swimming~\cite{Yao:2024:SwimmingVideos, Tang:2026:SwimChrono} and basketball~\cite{Lin:2023:Omnioculars, Chen:2023:iBall}.
Regarding emphasizing the audience's understanding of a competition, researchers in the visualization community have explored how to augment videos with data-driven overlays guided by natural language commentary (NLP), including \textit{VisCommentator}~\cite{Chen:2022:VisCommentator} for table tennis and \textit{Sporthesia}~\cite{Chen:2023:Sporthesia} for racket-based sports.
More generally, the concept of ``visualization in motion''~\cite{Yao:2022:VIM} applies to all cases of embedded visualizations in sports videos when entire visualizations move across the screen, such as those attached to athletes or sports equipment. It is inherently difficult to design and prototype such visualizations \cite{Yao:2024:SwimmingVideos}, even without initially considering storytelling elements, \revision{as they need to stay coupled with moving data referents, encode dynamically updating data, and remain suitable when the embedding context changes.}

Sports videos \revision{are particularly challenging for embedding visualizations because they are} inherently multimodal, combining video, speech, and textual elements. \revision{This diverse context adds difficulties to storytelling, as each channel supports different aspects}: video conveys raw visual content, text provides additional information, such as metadata and results, and commentary tracks events over time. However, these modalities remain loosely coordinated, resulting in fragmented narrative structures.
What remains underexplored is how to unify these elements to tell data-driven stories through visual narratives in sports videos,
\revision{shifting the authoring problem from simply embedding visualizations into video to coordinating data encodings, view changes, animated transitions, and race events over time.}
Professional sports broadcasts currently provide limited practical storytelling techniques: they direct the audience's attention by switching between different views, emphasizing key moments through close-up shots \cite{Tang:2022:SmartShots}, including audio commentary \cite{Chen:2022:VisCommentator}, and, for some sports such as swimming \cite{Yao:2024:SwimmingVideos, Tang:2026:SwimChrono}, injecting simple data text labels like speed to enable audiences to perceive information in an intuitive way.
However, these practices primarily rely on cinematic storytelling through camera work and scene editing, rather than on narratives that explicitly integrate data, views, and visual effects. Possible reasons include challenges in data preparation, which require both domain knowledge and processing skills, a lack of guidance for narrative visualizations in sports videos, and the absence of authoring tools, highlighting a clear research opportunity.

\revision{We investigate how to support authoring narrative visualizations in motion, using swimming as a case because of their rich dynamic data, practical embedded visualizations (\eg, speed labels), and billions of global Olympic audience reach \cite{IOC2024Audience}. 
Previous investigation shows that general audiences are interested in seeing more visualizations in swimming as they are very useful \cite{Yao:2024:SwimmingVideos}.}

To gather comprehensive tracking data rather than manual labeling, we developed \textbf{an automatic multimodal data preparation pipeline} that produces structured, visualization-ready data from three easy-to-obtain inputs: a single race video, relevant commentary audio, and a competition name.
To learn and be inspired by professional broadcasts on how to organize narrative attention, we conducted \textbf{an observational analysis} of broadcasts across swimming, basketball, and soccer.
To support content creators in authoring narrative visualizations in motion, we designed and implemented \textbf{a technology probe}~\cite{Hutchinson:2003:TechnologyProbes}, \SCe.
Grounded in our structured data and observed narrative patterns, \SCe\ frames authoring as coordinating views, transitions, data layers, and pacing over time, enabling creators to compose story-driven augmented race videos that explain a competition from start to finish rather than placing isolated overlays.
To assess the benefits and understand the challenges of authoring narrative visualizations in motion, we conducted \textbf{an empirical evaluation} with participants who are experienced in content creation and graphic design. Our results show that \SCe\ \revision{generally} enables effective narrative authoring and storytelling, with consistent authoring patterns, \revision{and remaining design challenges.
We end our paper with limitations, discussions and open research challenges for narrative visualization in motion.}

\section{Related Work}
We discuss here narrative and cinematic theory, which informs narrative techniques; visualization in motion, where our work lies; and tools for embedding visualizations in videos, related to our technology probe.

\subsection{Narrative Visualization and Data Videos}
Cohn's theory of visual narrative structure~\cite{Cohn:2013:VisualNarrativeStructure} noted that even simple image sequences can be interpreted differently, highlighting the role of narrative structure in organizing events into meaningful and communicable forms.
In visualization, narrative approaches similarly focus on structuring information to guide audience understanding. Segel and Heer~\cite{Segel:2010:NarrativeVisualization} proposed a design space, indicating that effective data stories rely on ordering, emphasis, and transitions to communicate meaning. 
Satyanarayan and Heer~\cite{Satyanarayan:2014:Ellipsis} further introduced \textit{Ellipsis} to support authoring such narratives through data-driven stories.
In data videos, prior work further shows how cinematic techniques support explanation. Amini \etal~\cite{Amini:2015:UnderstandingDataVideos} highlighted camera framing and transitions; 
Cao \etal~\cite{Cao:2020:NarrativeConstructs} identified narrative constructs such as emphasis and sequencing; 
Li \etal~\cite{Li:2023:GeoCamera} proposed \textit{GeoCamera} to show how camera movements (\eg, pan, zoom, follow) can serve different narrative purposes; 
and Shi \etal~\cite{Shi:2021:CommunicatingWithMotion} characterized how animation and transitions support storytelling.
More recently, Xu \etal~\cite{Xu:2022:CinematicOpening, Xu:2023:CinematicEndings} investigated how transitions and temporal structure shape narrative flow, and provided guidelines for creating openings and endings for data videos.
Together, these works suggest that narrative visualization in videos relies on coordinating camera movement, transitions, emphasis, and temporal structure.
We build on these insights in our technology probe by exposing camera movement, transitions, and animation as authoring operations for constructing narrative visualizations in motion.

\subsection{Visualization in Motion}
Yao \etal~\cite{Yao:2022:VIM} defined ``visualization in motion'' as visual data representations used in contexts that exhibit relative motion between the visualization and the viewer, demonstrating that people can read reliable information from visualizations at a fast speed and under irregular trajectories. 
Yao \etal~\cite{Yao:2024:SwimmingVideos} next explored embedding visualizations in motion in a real-world swimming video. They investigated audiences' data interests, proposed corresponding representations (\revision{we follow their dedicated visualization designs for swimming data in this work}), and implemented an authoring tool. 
Yao \etal~\cite{Yao:2025:VideoGame} then evaluated the user experience of visualizations in motion, showing that many trade-offs arise in design when reading a visualization is not the primary task. 
Researchers also examined mobile and wearable contexts. Islam \etal~\cite{Islam:2022:FitnessTracker} outlined a dedicated research agenda on visualizations in motion for fitness trackers, while Grioui \etal~\cite{Grioui:2024:Smartwatch, Grioui:2025:Smartwatch} conducted studies on the readability and design of micro-visualizations on smartwatches during movement.   
These works established the foundations of visualization in motion and explored questions related to perception and design.  In contrast, our work uses visualization in motion in a storytelling context, focusing on narrative authoring that unfolds over time.

\subsection{Tools for Embedding Visualizations in Videos}
Authoring visualizations embedded in or synchronized with video requires specific support beyond simple chart overlays.
\revision{Existing work has explored this problem for both general-purpose data-video authoring and sports visualization.
General-purpose data-video tools, such as \textit{DataClips}~\cite{Amini:2017:DataClips}, \textit{InfoMotion}~\cite{Wang:2021:InfoMotion}, and \textit{WonderFlow}~\cite{Wang:2025:WonderFlow}, support data-video creation through templates, structural analysis, or narration-driven workflows. Other tools like \textit{Gemini}~\cite{Kim:2021:Gemini, Kim:2021:Gemini2}, \textit{AutoClips}~\cite{Shi:2021:AutoClips}, and \textit{Roslingifier}~\cite{Shin:2023:Roslingifier} focus on animated transitions and automation, while more recent systems such as \textit{Data Player}~\cite{Shen:2024:DataPlayer}, \textit{Data Playwright}~\cite{Shen:2025:DataPlaywright}, and \textit{Live Charts}~\cite{Ying:2025:LiveCharts} incorporate intelligent assistance. 
These tools advance data-video authoring, but they primarily focus on presentation-style videos, where the visualization is the main content, rather than on embedding visualizations into existing video footage with moving entities, camera changes, and unfolding events that are the main focus.

Sports visualization research has explored coupling visualizations with dynamic sports data in videos.
For example, Stein \etal~\cite{Stein:2018:BringItToThePitch} linked soccer videos with movement data to support tactic analysis. 
Lin \etal~\cite{Lin:2023:BallCourt, Lin:2024:VIRD} reflected on visualization research with sports experts and further investigated immersive VR video analysis for high-performance badminton coaching.
Researchers also worked on helping general audiences understand sports events. 
Chen \etal~\cite{Chen:2022:VisCommentator} proposed \textit{VisCommentator}, embedding static and animated visualizations on static table-tennis video frames to supplement commentary. They further proposed \textit{Sporthesia}~\cite{Chen:2023:Sporthesia}, which extracted commentary data via NLP and overlaid corresponding visual designs again on static video frames.
Besides, Lee \etal~\cite{Lee:2025:Sportify} proposed \textit{Sportify}, a visual question-answering system that embeds visualizations into basketball videos to help fans understand tactics.

Apart from overlaying visualizations statically in videos, recent work has enabled visualizations to move with athletes in sports videos, which is closer to our work.
Lin \etal~\cite{Lin:2023:Omnioculars} designed \textit{Omnioculars}, which attaches visualizations to moving players in a simulated basketball game to improve the watching experience.
Chen \etal~\cite{Chen:2023:iBall} proposed \textit{iBall}, which embeds gaze-moderated visualizations in basketball videos and adapts the display according to the viewer's gaze. 
Subsequently, Lin \etal~\cite{Lin:2025:SportsBuddy} proposed \textit{SportsBuddy} to support authoring highlight basketball clips through embedded visualizations and player tracking.
More closely aligned with our swimming context, Yao \etal~\cite{Yao:2024:SwimmingVideos} proposed \textit{SwimFlow}, which enabled users to design, edit, and embed visualizations in motion from a single view within swimming videos.
Tang \etal~\cite{Tang:2026:SwimChrono} examined when and for how long visualizations should appear in swimming videos and integrated their findings into \textit{SwimChrono}.

Together, prior work has advanced data-video authoring, sports explanations, and the dynamic embedding of visualizations in sports videos. This line of work is closest to ours but mainly focuses on viewer engagement, highlighting, visualization embedding, or temporal arrangements. Our work addresses a different authoring problem: full-race storytelling for general audiences.
We coordinate the visualization, moving entities, race events, views, and transitions over time, and shape an entire competition into a coherent video narrative.
}

\section{An Automatic Multimodal Data Preparation Pipeline}
\label{sec:pipeline}
  
\begin{figure*}[t]
    \centering
    \includegraphics[width=\textwidth]{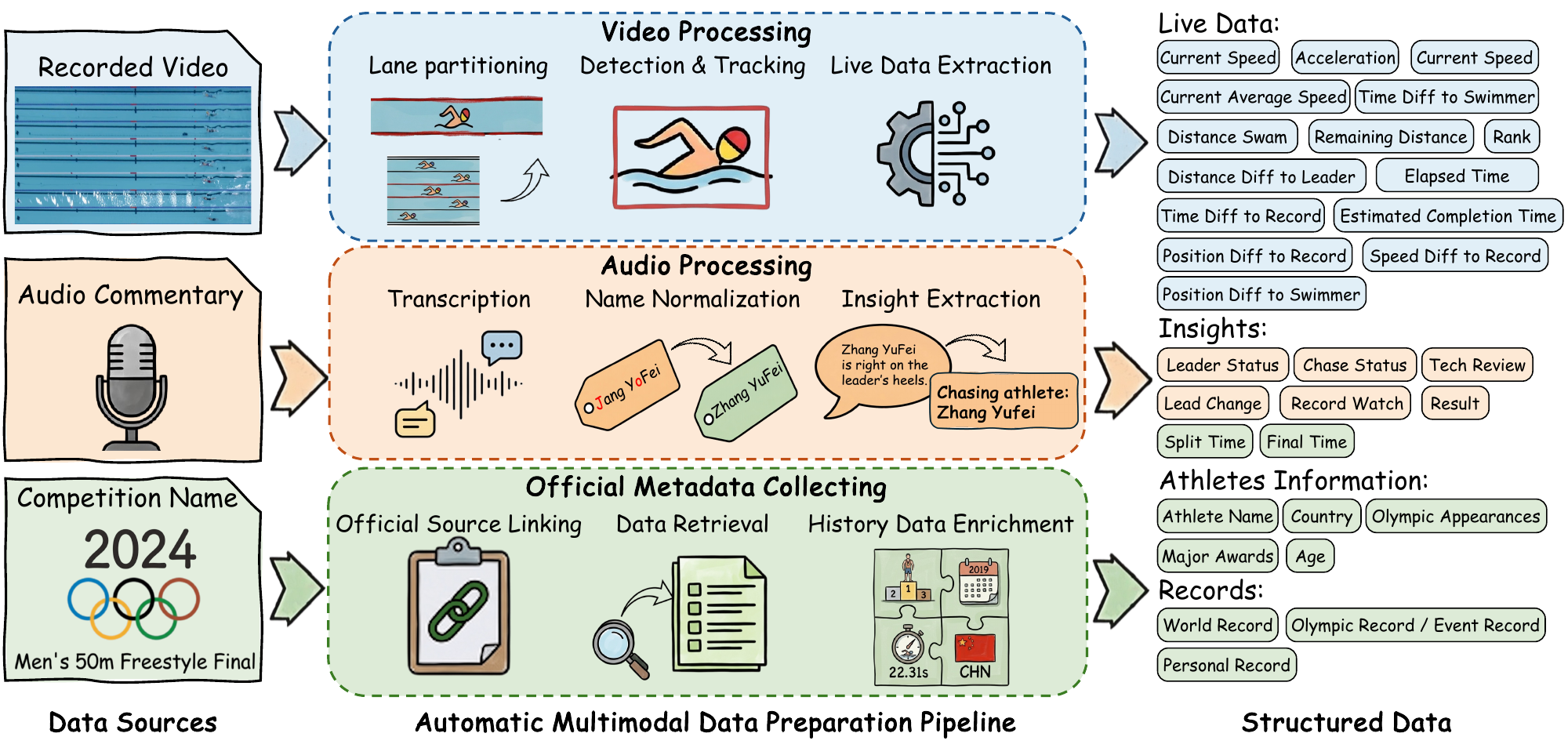}
    \caption{Our automatic multimodal data preparation pipeline. Starting from a recorded race video, commentary audio, and a competition name, the pipeline produces unified structured visualization data, including live data, event insights, athlete information, and records.}
    \label{fig:data_pipeline}
\end{figure*}

Our description begins with data preparation, which used to be the most labor-intensive step in our prior work on embedding visualizations into videos. Previously, we had to manually annotate data, such as athletes' positions and event timestamps, or extract it using custom tools before aligning it across sources. This process was time-consuming, difficult to reproduce, and hard to transfer across races.  
To address these limitations, we designed an automatic multimodal data preparation pipeline (\autoref{fig:data_pipeline}). Our pipeline integrates video analysis, audio understanding, and metadata retrieval into a unified workflow, taking three easy-to-obtain inputs: a recorded race video, commentary audio, and a competition name. From these inputs, our pipeline produces structured data organized into four temporally aligned groups: \textit{Live Data}, \textit{Event Insights}, \textit{Athlete Information}, and \textit{Records}, ready for visualization.
We describe our pipeline below, from data sources to video and audio processing and metadata retrieval.
We release source code of our pipeline (see supp.\ material section for the pointer).

\subsection{Data Sources}
To keep the input requirements minimal, we restrict our pipeline to three simple data sources:

\vspace{1pt}
\noindent \textbf{Reordered Video:} 
We used our own videos recorded under authorization at the 2024 Paris Olympic swimming races, combined from left- and right-separated recordings by computer vision processing scripts from Yao \etal~\cite{Yao:2024:SwimmingVideos}. 
The resulting videos provide a bird's-eye view (see the example in ``Recorded Video'' under ``Data Source'' in \autoref{fig:data_pipeline}) with full context of the swimming pool and serve as input to our pipeline.

\vspace{1pt}
\noindent \textbf{Audio Commentary:} 
We collected commentary audio for the corresponding race from official broadcasts when they are available. For races without official commentary tracks, we use commentary audio from major media outlets, such as NBC Sports. This source captures event-level descriptions and insights of the race progression that are often not explicit in the visual signal alone.

\vspace{1pt}
\noindent \textbf{Competition Name:} We used the competition name as a single competition identifier, such as ``\textit{Paris 2024 Men's 50m Freestyle Final}'', as the entry point for metadata retrieval. This compact input is sufficient to resolve the official event pages and retrieve both event-level and athlete-level information without manual name collection.

We use these inputs as they are widely available and provide complementary information: video for motion analysis, commentary for event insights, and competition name for metadata. Together, they enable data preparation without manual annotation.

\subsection{Video Processing}
Our goal for the video processing pipeline is to extract fine-grained, time-varying performance data from race videos with minimal human intervention.
Existing systems often rely on manual labels or on supervised models trained on a specific dataset.
In contrast, we leverage the structural regularity of swimming videos with prompt-based segmentation and tracking, including three steps: lane-based spatial partitioning, temporal segmentation and tracking, and live data extraction.

\vspace{2pt} 
\noindent \textbf{Lane Partitioning:}
In the bird’s-eye view, swimming lanes appear as parallel and non-overlapping. As swimmers remain within their lanes, each can be treated as an independent motion channel. We divide the frame into fixed lane regions and process them separately, simplifying segmentation and reducing cross-lane interference.

\vspace{2pt} 
\noindent \textbf{Detection and Tracking:}
Within each lane, we use SAM3~\cite{carion2025sam3segmentconcepts}, a foundation model that supports prompt-based object localization and temporal tracking of the swimmer.
As each region contains at most one swimmer after lane partitioning, a simple text prompt is sufficient to initialize the target, which is then propagated across frames to recover positions.
This reduces manual labeling and race-specific detector training, improving the adaptability of our pipeline across races.

\vspace{2pt} 
\noindent \textbf{Live Data Extraction:}
From the segmented swimmer masks and tracked positions, we derive lane-wise trajectories and compute time-varying live data.
Per frame, our pipeline records each swimmer's horizontal position within their lane 
and computes instantaneous speed, acceleration, and cumulative distance. By comparing positions across lanes, our pipeline derives frame-level rankings and pairwise gaps.
These per-frame metrics are smoothed over a short temporal window to reduce noise from segmentation jitter.

The result of video processing is a structured live data stream aligned with the race timeline. This stream serves as the dynamic foundation of the \textit{live data} layers in our technology probe and can be reused across races without lane-by-lane manual annotation.

\subsection{Audio Processing}
\label{sec:audioprocessing}

Video-derived signals capture continuous motion but miss the narrative interpretations that commentators provide to audiences, such as leader introductions, close-ups, and takeovers. We therefore process commentary audio to extract structured event insights through transcription, name normalization, and insight extraction, to supplement the video-derived data.
\revision{For name normalization and insight extraction, we use an LLM (GPT-5.4) with fixed prompts. We provide our prompts, output schemas, and validation scripts for both steps in the supp.\ material.}

\vspace{2pt} 
\noindent \textbf{Transcription:}
We use WhisperX~\cite{bain2022whisperx} to automatically transcribe the commentary audio. It segments speech into words and assigns each word a start and end timestamp. This word-level temporal resolution is finer than sentence-level transcription and enables precise alignment between commentary and race events in later stages.

\vspace{2pt} 
\noindent \textbf{Name Normalization:}
Automatic transcription often produces inconsistent spelling of athlete names due to accents, languages, and broadcast conditions.
\revision{Using the official athlete roster retrieved by our metadata pipeline, the LLM normalizes mentioned names to official athlete names.}
This step ensures that commentary reliably links to the correct swimmers and is later synchronized with lane-level race data.

\begin{figure}[t]
  \centering
  \includegraphics[width=\columnwidth]{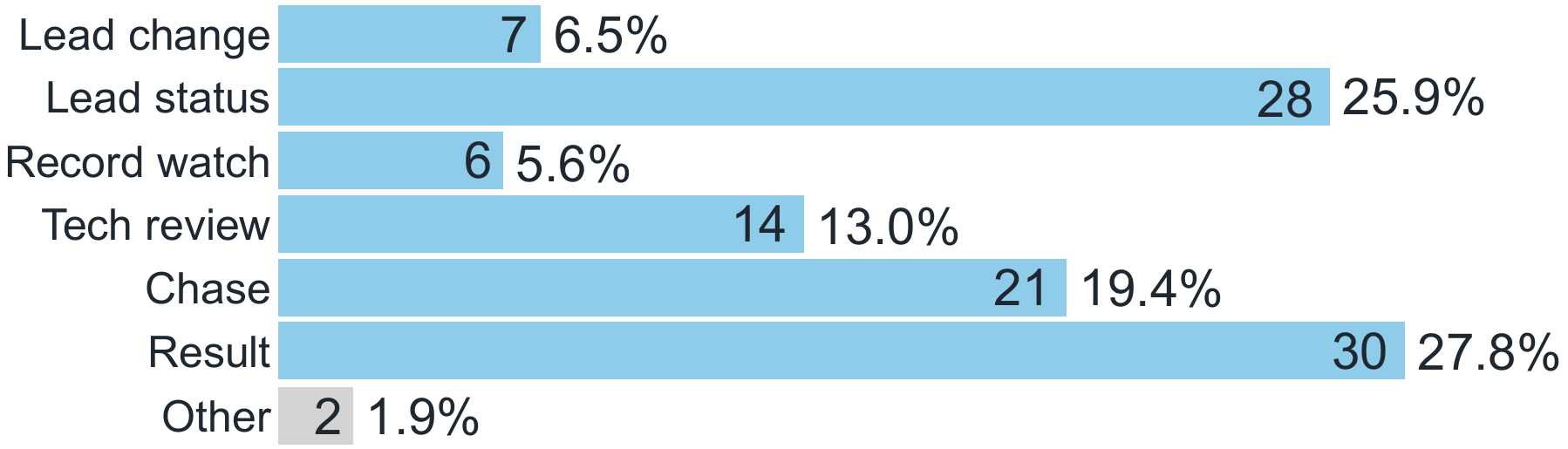}
  \caption{Distribution of coded key moments across the proposed insight types. 108 commentary units were coded as key moments across 15 races, and 106 of them (98.15\%) were covered by our insight categories.}
  \label{fig:insight-distribution}
\end{figure}

\vspace{2pt} 
\noindent \textbf{Insight Extraction:}
\revision{The LLM identifies any applicable predefined insight types for each normalized segment.}
These include \textit{lead change} (a change in the current leader), \textit{lead status} (descriptions of who is leading or maintaining an advantage), \textit{record watch} (commentary about record pace or record-breaking possibility), \textit{tech review} (technical commentary on turns, underwater phases, or stroke execution), \textit{chase} (close pursuit between swimmers), and \textit{result} (final outcome or placement).
While some of these events can be approximated from tracking data, they are primarily extracted from commentary as event-driven interpretations that reflect how the race is narrated and perceived. 
These insights capture moments that commentators emphasize to describe the race's progression and highlight key moments (\autoref{fig:insight-distribution}). 
We validated our insight design using commentary from 15 swimming races across five tournament series: the Olympic Games Paris 2024, the Olympic Games Tokyo 2020, the World Aquatics Championships Doha 2024, the World Aquatics Championships Singapore 2025, and the Swimming World Cup 2025. The dataset covers both men’s and women’s events across freestyle, butterfly, backstroke, breaststroke, and individual medley, with distances from \m{50} to \m{800}. We segmented the commentary into short units and coded each as either a key moment or a non-key contextual comment. Only key-moment units were used for validation. Of 136 units, 108 were key moments, and 106 were covered by our proposed insight types, giving a coverage rate of 98.15\%. This suggests that our insight design covers most commentator-emphasized race events and is reliable enough for the \textit{insights} layer in our system.

The output of audio processing is a sequence of timestamped event insights.
These insights do not reproduce commentary but abstract key narrative cues, often decomposing a single utterance into multiple structured insights. 
These insights are not meant to replace video-derived signals but to complement them, serving as the foundation of the \textit{insights} layer in our technology probe. 

\vspace{-4pt}
\subsection{Official Metadata Collection}

Traditional metadata collection often requires manual browsing of result pages, lane sheets, and athlete profiles.
This process is tedious and can easily introduce inconsistencies across races.
To reduce this burden, we built an AI-assisted metadata retrieval pipeline.
Given a single competition name, the pipeline resolves the target event, identifies the relevant official sources, extracts event-level metadata, and then enriches the result with athlete-level background and record data.
The pipeline has three stages: competition resolution and source linking, official event data retrieval, and athlete history and record enrichment.

\vspace{2pt} 
\noindent \textbf{Official Source Linking:}
Given a competition name such as ``\textit{Paris 2024 Men's 50m Freestyle Final}'', the pipeline parses the name and locates the corresponding official competition and event pages. This stage focuses on source resolution rather than field extraction: it maps a compact textual identifier to the correct Olympics and World Aquatics pages and establishes the source set for subsequent retrieval. AI at this stage handles naming variations and keeps the input compact.

\vspace{2pt} 
\noindent \textbf{Data Retrieval:}
After the target sources are resolved, the pipeline performs structured extraction from the linked event pages.
This stage retrieves event-level fields, including athlete identities, lane assignments, country information, result-related fields, and official record markers such as the world and Olympic records.
The extracted fields are then normalized into a unified schema that can be used directly by the visualization system.
This step provides the structured event context for the target race.
In addition, official split times and final times are forwarded to the \textit{Insights} group (\autoref{sec:audioprocessing}), where they serve as timing-based narrative anchors alongside commentary-derived insights.

\vspace{2pt} 
\noindent \textbf{History Data Enrichment:}
Using the athlete identities from the previous stage, the pipeline retrieves athlete-level information from official sources. These data include major awards, Olympic appearances, and personal records, completing event-level metadata with athlete context useful for comparison-based visuals and narrative storytelling.

We restrict our pipeline to official sources, which provide stable event-specific metadata and clearer provenance than open web search.
In this design, AI serves as the automation layer rather than a factual source: it parses the competition name, coordinates retrieval, and structures the output, while the facts remain grounded in official competition, event, and athlete pages.
Because the same pipeline can be reused for new races with only a competition name as input, it also improves reproducibility.
The final outputs are written into two structured groups used later in our technology probe: \textit{Athlete Information} and \textit{Records}.


\vspace{-4pt}
\section{From Sports Broadcasts to Narrative Supports}
\label{sec:frombroadcasts}
To understand how professional sports broadcasts narrate sports events and to draw inspiration from professional practices, we analyzed three widely watched broadcasts: the women's \m{400} freestyle swimming final at the 2024 Paris Olympics (\mins{4}), an NBA basketball game between the San Antonio Spurs and the Charlotte Hornets (\mins{85}), and a La~Liga soccer match between Villarreal and Real Madrid (\mins{98}).
We selected these broadcasts because they represent mature production practices across three widely followed sports and offer contrasting narrative conditions, from the lane-based swimming progression to the fast-paced basketball back-and-forth play to the broader soccer spatial organization. Such diversity allows us to derive ideas for techniques that can inform narrative authoring in swimming videos.
We focus on three aspects closely related to narrative authoring: camera views, transitions between views, and the treatment of key moments.
Our observations show how professional broadcasts shift narrative focus and emphasize key moments. Rather than treating broadcasts as templates to reproduce, we use them as inspiration to derive narrative supports for content creators in our probe.
We report our concrete observations and counts from three broadcasts below (original data available in supp.\ materials). Our findings are not intended to generalize to all broadcasts per sport, but to illustrate patterns that can inform narrative authoring.

\vspace{-4pt}
\subsection{Transitions in Sports Broadcasts}
Since our goal is to identify patterns relevant to narrative authoring rather than build a detailed typology of broadcast cinematography, we categorize views by the level of detail shown to audiences: overview and focus-on.
Across three broadcasts, we manually annotated 292 transitions and their timestamps (\percentagebar{0.17} 50/292 from swimming, \percentagebar{0.37} 108/292 from basketball, and \percentagebar{0.46} 134/292 from soccer). Each transition starts when the broadcast moves from overview to focus-on and ends when it returns to overview. We report our findings along three dimensions: transition types, their usage, and functional moments.

\vspace{2pt}
\noindent\textbf{Transition Types and Their Prevalence:}
We identify three transition types: \textit{hardcut}, \textit{wipe}, and \textit{zoom-in}.
\textit{Hardcut} denotes a direct switch between shots.
\textit{Wipe} replaces one shot with another through a moving boundary or animated broadcast graphic, such as a moving logo or overlay that conceals the cut.
\textit{Zoom-in} indicates a camera movement or reframing that brings a specific target closer into view.
Across all collected broadcasts, \textit{hardcut} was the most frequent transition type (\percentagebar{0.747} 218/292), followed by \textit{wipe} (\percentagebar{0.226} 66/292) and \textit{zoom-in} (\percentagebar{0.027} 8/292). 
This pattern varied by sport for our analyzed broadcasts: swimming used \percentagebar{0.84} 42/50 \textit{hardcut}, \percentagebar{0.16} 8/50 \textit{zoom-in}, and no \textit{wipe}; basketball used \percentagebar{0.769} 83/108 \textit{hardcut}, \percentagebar{0.231} 25/108 \textit{wipe}, and no \textit{zoom-in}; and soccer used \percentagebar{0.694} 93/134 \textit{hardcut}, \percentagebar{0.306} 41/134 \textit{wipe}, and no \textit{zoom-in}.

\vspace{2pt}
\noindent\textbf{Transition Use as Broadcast Status Cues:}
In these broadcasts, we observed distinct transition patterns depending on whether the footage was live or a replay.
In basketball and soccer, all \textit{hardcuts} occurred in live footage, while almost all \textit{wipes} (63/66) appeared in replays, suggesting that transition types may cue audiences to distinguish ongoing action from retrospective review.
Swimming showed no such distinction: all 50 transitions occurred within live footage, with no replay sequences observed. Such absence indicates that this swimming broadcast relied on within-live devices such as \textit{zoom-in} to guide attention.

\vspace{2pt}
\noindent\textbf{Transition at Key Moments:}
Transitions were closely tied to key moments, though their prevalence varied across sports.
In basketball, all transitions (108/108) were associated with at least one key moment, most commonly 73/108 scoring, followed by 9/108 free throws, and 7/108 falls.
In soccer, half of the collected transitions (67/134) aligned with key events,  including 23/67 falls, 9/67 out-of-bounds, 7/67 shots, and 5/67 yellow cards.
Swimming showed a distinct pattern: all \textit{zoom-ins} (8/8) coincided with the emergence of the race leader, while only 5/42 \textit{hardcuts} aligned with key events.
This suggests that \textit{zoom-ins} in the swimming broadcast were not used as general-purpose transitions, but as focused narrative devices to emphasize leader-related moments.

\vspace{-4pt}
\subsection{Narrative Structures in Broadcast Videos}
Across the three broadcasts, we observed consistent narrative structures in how views, transitions, and key moments guide audience attention over time. We summarize these structures along three dimensions: view types, connections between views, and emphasis methods:

\vspace{2pt}
\noindent\textbf{Different Views:}
Although our collected entities do not quantify view distribution, we consistently observed a recurrent alternation between two complementary viewing modes across all three broadcasts: \textit{overview} and \textit{focus-on}.
\textit{Overview} maintains the competition's overall state, preserving spatial context, progression, and relations among athletes, while \textit{focus-on} directs attention to a specific athlete, action, or key event, such as a leading swimmer, a scorer, or a foul.
This recurrence shift suggests it may reflect a common narrative operation in broadcast practice and indicates that narrative authoring in motion should support both global and detailed views, as well as creators' decisions about when to move between them.

\vspace{2pt}
\noindent\textbf{View Connection:}
\textit{Hardcut} was most often used during live play to connect views, while \textit{Wipe} appeared mainly around replay or broadcast-graphic segments in the two broadcasts of basketball and soccer, signaling a shift from live progression to retrospective review.
In the analyzed swimming broadcast, \textit{zoom-in} served as a dedicated emphasis method, directing audiences' attention to a specific swimmer, typically the leader or another athlete of immediate narrative interest. 
These findings show that transitions do more than connect views: different transition types signal how audience attention is being redirected.

\vspace{2pt}
\noindent\textbf{View Emphasis:}
Shifts between \textit{overview} and \textit{focus-on} often occurred around key moments, such as scoring, fouls, stoppages, and replays in basketball and soccer, and start (diving), turns, emerging leaders, overtakes, and record-breaking in swimming.
In swimming, we observed \textit{zoom-ins} consistently co-occurred with leader-related events, suggesting that certain transitions were reserved for narratively important progression.
This pattern shows that narrative authoring could benefit from event-driven emphasis, enabling creators to identify key moments and structure the timing and intensity of narrative attention.

In summary, based on our analyzed broadcasts, audiences' attention is guided through three elements: alternating between two views (\textit{overview} and \textit{focus-on}), using transitions such as \textit{hardcut}, \textit{wipe}, and \textit{zoom-in} for view shifting, and aligning these transitions with key moments. Our observations suggest that, in addition to data, authoring narrative visualization in motion also requires support for views, transitions, and events to unfold the race over time.

\section{Towards Authoring Narrative Vis in Motion:\\A Technology Probe --- \SCe} 
\label{sec:swimcomposer}

\begin{figure*}[t]
  \centering
  \includegraphics[width=\textwidth]{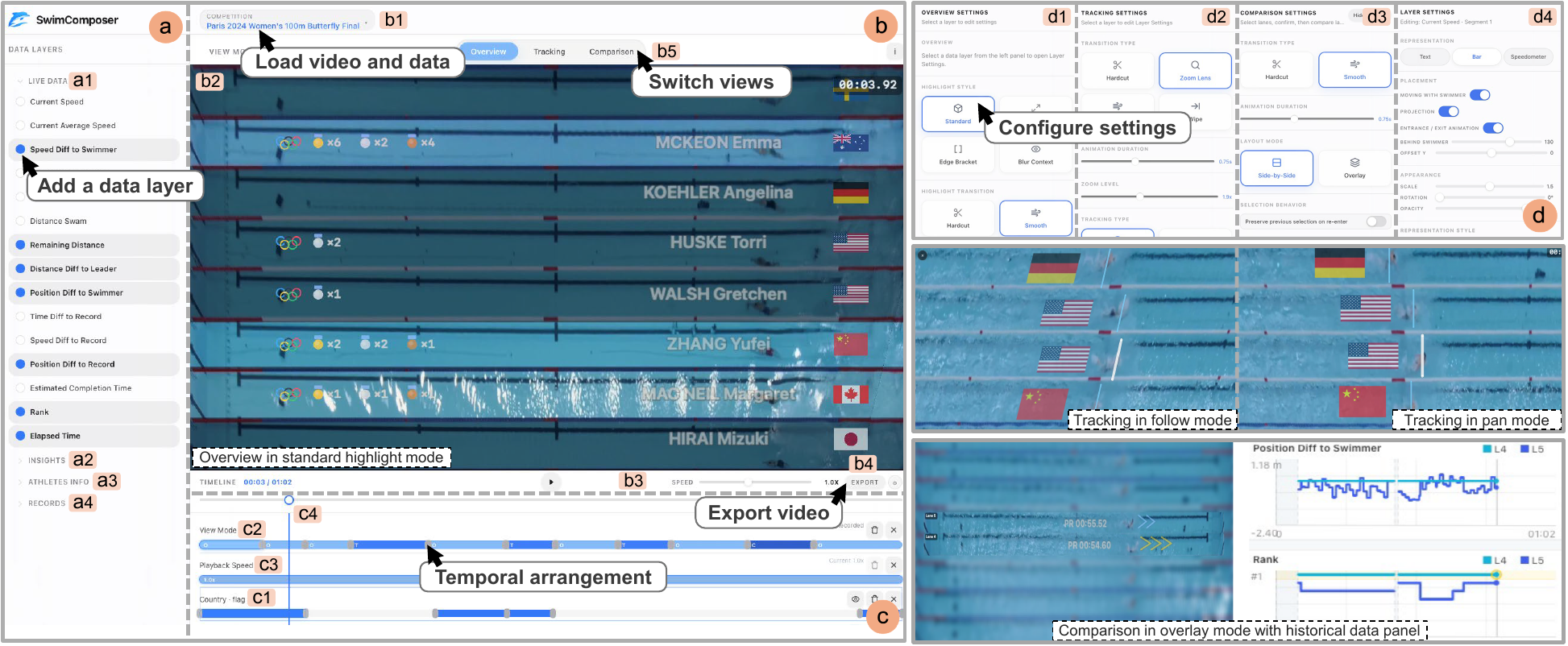}
  \caption{The \textit{SwimComposer} workspace, interactions, and example illustrations of different views. 
  \orangecircle{a}~\texttt{Layer Library} organizes race information into four categories: \panellabel{a1} \textit{Live Data}, \panellabel{a2} \textit{Insights}, \panellabel{a3} \textit{Athlete Information}, and \panellabel{a4} \textit{Records}. 
  \orangecircle{b}~\texttt{Narrative Viewing Panel} has \panellabel{b1} a video selector and \panellabel{b3} play controls, \panellabel{b2} displays the race video with embedded data layers and provides \panellabel{b5} three view selectors: \textit{Overview}, \textit{Tracking}, and \textit{Comparison}. It additionally has \panellabel{b4} an export button for exporting final designs. 
  \orangecircle{c}~\texttt{Narrative Timelines} arrange the temporal composition of all narrative elements: \panellabel{c1} \textit{Layer Segments}, \panellabel{c2} \textit{View Mode Segments}, \panellabel{c3} \textit{Playback Speed}, and \panellabel{c4} \textit{Preview}. 
  \orangecircle{d}~\texttt{Configuration Panels} provide \panellabel{d1} \textit{Overview Settings} including highlight modes and transitions, \panellabel{d2} \textit{Tracking Settings} including tracking modes, zoom level, transitions, and animation speed, \panellabel{d3} \textit{Comparison Settings} including comparison mode, transitions, and animation speed, and \panellabel{d4} \textit{Layer Settings} including representation selection, placement, and design parameters.
  Users first \solidlabel{load video and data}. They can then \solidlabel{add a data layer} or \solidlabel{switch views}. They \solidlabel{configure settings} for narrative elements, and adjust their \solidlabel{temporal arrangement}. Once complete, they \solidlabel{export video}. 
  \dottedlabel{Overview in standard highlight mode} illustrates \textit{highlighting} swimmers in lanes 2, 6, and 7.
  \dottedlabel{Tracking in follow mode} and \dottedlabel{Tracking in pan mode} show two \textit{tracking} modes.
  \dottedlabel{Comparison in overlay mode with historical data panel} demonstrates the \textit{overlay} effect, accompanied by selected data dashboards.
  }
  \label{fig:workspace}
\end{figure*}

Based on the structured data generated by our automated pipeline and the patterns observed in our broadcast analysis, we implemented \SCe, a technology probe~\cite{Hutchinson:2003:TechnologyProbes} for authoring narrative visualizations in motion in swimming videos. 
Apart from supporting the embedding and direct manipulation of visual representations under motion, 
\SCe\ enables authors to construct narratives by embedding visualizations in motion in videos, coordinating between global and detailed views, tracking and comparing across swimmers, highlighting key moments, and transitioning across narrative states. 
\SCe\ (\autoref{fig:workspace}) has four main parts (see \autoref{sec:SwimComposerWorkspace}).
%
%
The videos we used in \SCe\ are our own recordings authorized for the 2024 Paris Olympics, combined from left- and right-separated recording footage using the processing methods and scripts released by Yao \etal~\cite{Yao:2024:SwimmingVideos}.
%
%
\revision{Most visualization designs included in \SCe\ are from the dedicated swimming data representations proposed by Yao \etal~\cite{Yao:2024:SwimmingVideos}. For example, lines indicate world and Olympic records, while bar and speedometer/gauge charts represent speed-related information. For data not included in Yao \etal's designs, we use text as the default representation to keep the focus of our work on narrative authoring rather than on proposing new graphical encodings.}
%
%
We developed the \SCe\ interface using web technologies, including JavaScript, HTML, and CSS.
The graphical implementation and motion rendering were implemented using a GPU-accelerated PixiJS canvas overlaid on the HTML video, enabling smooth, real-time rendering.
The input to \SCe\ includes the structured race data generated by our pipeline described in \autoref{sec:pipeline}. 
At load time, \SCe\ ingests the pipeline outputs and instantiates the default representation for each data item. 
We next explain our design considerations of \SCe, followed by a description of its user interface.  
\SCe\ can be accessed at \href{http://43.163.231.237/}{\texttt{43\discretionary{}{.}{.}163\discretionary{}{.}{.}231\discretionary{}{.}{.}237\discretionary{/}{}{/}}}.

\subsection{Design Considerations}
Our broadcast analysis in \autoref{sec:frombroadcasts} suggests that authoring narrative visualizations in motion requires different views to guide attention, transitions between views, and emphasizing key moments to unfold the race.
We derive four design considerations for our technology probe:

\noindent \roundedDClabel{DC1} \textbf{Accessible Data and Editable Representations:} 
A major difference between our work and professional broadcasts is that our narratives are driven not only by race progression but also explicitly by data. 
While our pipeline from ~\autoref{sec:pipeline} produces rich structured data, these data must be accessible and editable within the technology probe to support narrative construction. This enables users to select, design, embed, and adapt data representations in video to better explain the race.
We organize the available data following the categories derived from our pipeline: \textit{live data}, \textit{insights}, \textit{athlete info}, and \textit{records}.    

\noindent \roundedDClabel{DC2} \textbf{Different Views:}
Our broadcast analysis showed that narratives across all three broadcasts were organized through a consistent alternation between global and detailed views.
The global view maintains the race's overall state, while the detailed view focuses on specific athletes or key moments.
In swimming, however, race understanding additionally relies on comparing athletes across lanes to explain distance differences, lead changes, and turns. 
We therefore include three views in our technology probe: \textit{overview} for all swimmers, \textit{tracking} for individual athletes, and \textit{comparison} for relating multiple swimmers.

\noindent \roundedDClabel{DC3} \textbf{Transitions between Views:} 
According to our broadcast  observations in \autoref{sec:frombroadcasts}, transitions, including \textit{hardcut}, \textit{zoom-in}, and \textit{wipe}, are commonly used to shift views or emphasize moments. 
To provide greater flexibility, we also include \textit{dissolve}, drawn from cinema to convey temporal continuity, and animated pixel-level correspondences, which we call \textit{smooth}, to support gradual graphical change.
Transition choices depend on view characteristics. For example, an \textit{overview} can transition to \textit{tracking} via \textit{zoom-in}, as the focus shifts from global context to a single swimmer. This is less suitable for \textit{comparison}, which requires more space to assess relative distances and lead changes.
We thus provide \textit{hardcut} for both \textit{tracking} and \textit{comparison}; \textit{zoom-in}, \textit{wipe}, and \textit{dissolve} for \textit{tracking}, and \textit{smooth} transitions for \textit{comparison}.

\noindent \roundedDClabel{DC4} \textbf{Flexible Temporal Arrangement:} 
Narrative explanation in sports videos depends on timing at multiple levels. 
For data, creators need to determine not only what and how to show it,  but also when and for how long. 
For view transitions, they need to decide when to shift views and set an appropriate transition pace to maintain perceptual comfort. 
For the video, creators require standard play controls such as fast-forward and slow-motion.
We thus include a flexible temporal arrangement at three levels in our technology probe:
\textit{animation speeds} for transition pace; 
\textit{timelines} for specifying and adjusting the timing and duration of each data element and view; 
and \textit{playing speed} controls for adjusting the speed of both the video and embedded visualizations.

\vspace{-4pt}
\subsection{\SCe\ Workspace}
\label{sec:SwimComposerWorkspace}
The \SCe\ workspace includes four parts: a \texttt{layer library} \orangecircle{a} that organizes our structured data categories, a \texttt{narrative viewing panel} \orangecircle{b} with video selection, play controls, and view mode selectors, a set of \texttt{narrative timelines} \orangecircle{c} for temporal composition of all narrative elements, and a set of \texttt{configuration panels} \orangecircle{d} that provide dedicated controls for each narrative element.
Together, these components support creators in deciding \textit{what data, when and how} (\revision{most designs follow Yao \etal~\cite{Yao:2024:SwimmingVideos}; otherwise, we use text by default}) to present, \textit{which view} to adopt, \textit{when and how to transition} between views, and \textit{when and how} each element appears and disappears. Concrete illustrations are in \autoref{fig:workspace}.

\vspace{2pt}
\noindent\textbf{Layer Library (\autoref{fig:workspace}\orangecircle{a}):}
To support data-driven authoring (\roundedDClabel{DC1}), the layer library organizes all available race data generated from our pipeline into four categories below and exposes them as editable layers:

\noindent \panellabel{a1} \textit{Live Data} layers represent continuously evolving performance metrics computed from the race video, including speed, acceleration, distance swum, current ranking, and gap relative to other swimmers or records.
Each metric can be displayed using different visual representations. Creators can choose and adjust design parameters in the layer configuration panel (see \textbf{Configuration Panels} for details).

\noindent \panellabel{a2} \textit{Insights} layers surface key narrative moments automatically extracted by our pipeline from commentary audio and official race data.
These include lead status, tech review, chase status, split times, and race results.
Rather than requiring creators to manually identify which moments deserve emphasis, insights provide ready-to-use narrative anchors (\roundedDClabel{DC1}) that can be placed on the timeline to structure the story.

\noindent \panellabel{a3} \textit{Athlete Information} layers provide background and identity context about each swimmer, including name, nationality, age, major achievements, and Olympic history.
These layers help establish who is competing and why a particular swimmer matters, supporting the introductory or contextual elements of the narrative.

\noindent \panellabel{a4} \textit{Records} layers display reference benchmarks: world/Olympic records and personal bests.
Shown alongside live data, they help audiences immediately see whether a swimmer is on pace to break a record or how the current race compares to historical performances.

\vspace{2pt}
\noindent\textbf{Narrative Viewing Panel (\autoref{fig:workspace}\orangecircle{b}):}
This panel allows creators to preview the race video with embedded data layers.
It includes \panellabel{b1} video selection, \panellabel{b2} a video canvas, \panellabel{b3} playback controls (play/pause and speed slider), \panellabel{b4} an export button (see \textbf{Export} for details), and \panellabel{b5} view mode selectors (\ie, \textit{Overview}, \textit{Tracking}, and \textit{Comparison}; see \textbf{Configuration Panels} for details) to support switching between narrative views (\roundedDClabel{DC2}). Creators can switch views at any point, with each change recorded as a timeline segment, making view shifts an explicit part of the authored narrative. 

\vspace{2pt}
\noindent \textbf{Narrative Timelines (\autoref{fig:workspace}\orangecircle{c}):}
The set of timelines supports temporal composition of all narrative elements, including data layers, views, and transitions (\roundedDClabel{DC3}, \roundedDClabel{DC4}), where all elements are represented as editable segments for arrangement and adjustment:

\noindent \panellabel{c1} \textit{Layer Segments:}
Each active layer appears as a timeline segment, supporting creators to control its visibility and split it into multiple intervals.
This enables information to appear at narratively appropriate moments, rather than remain statically visible throughout the video.

\noindent \panellabel{c2} \textit{View Mode Segments:}
View mode changes are also represented as timeline segments, thereby structuring view shifts into temporal order.
For example, a creator may begin a race with an \textit{overview}, switch to \textit{tracking} to follow the leader in the middle, and then move to \textit{comparison} to highlight tiny differences in the final meters.

\noindent \panellabel{c3} \textit{Playback Speed:}
The timeline allows creators to assign different playback speeds (via speed slider in the narrative viewing panel) to different intervals.
This supports pacing control by slowing key moments for emphasis and speeding up less eventful segments.

\noindent \panellabel{c4} \textit{Preview:}
The video canvas updates in real time as creators scrub through or play the timeline, rendering all active layers, the current viewing mode, and any transitions at the corresponding point in time.
This immediate visual feedback allows creators to evaluate their authoring decisions in context rather than working from an abstract script.

\vspace{2pt}
\noindent \textbf{Configuration Panels (\autoref{fig:workspace}\orangecircle{d}):}
The right side of the workspace holds four configuration panels: \textit{Overview settings}, \textit{Tracking settings}, \textit{Comparison settings}, and \textit{Layer settings}:

\noindent \panellabel{d1} \textit{Overview Settings} control how swimmers are highlighted within the global view, offering four styles with different narrative effects. 
\textit{Standard} softly dims the surrounding lanes while keeping the full race context readable. 
\textit{Spotlight} darkens the area outside the selected swimmer more strongly to concentrate the attention. 
\textit{Edge Bracket} emphasizes lane boundaries with bracket-like marks near the pool edges, supporting moments such as turns or finishes. 
\textit{Blur Context} blurs the surrounding region while keeping the selected swimmer sharp.
We provide \textit{hardcut} and \textit{smooth} transitions for highlighting.

\noindent \panellabel{d2} \textit{Tracking Settings} control how attention follows a selected swimmer over time, including two modes, with a zoom slider provided for determining the tracking scope.
\textit{Follow} keeps the selected swimmer centered, simulating a side-view tracking shot (\revision{see \autoref{fig:workspace}: \dottedlabel{Tracking in follow mode}}), as if the camera were moving with the swimmer. 
\revision{We used homography transformations to implement the \textit{follow} function, enabling flexible simulation from different viewpoints of our videos. The view simulation applies to embedded visualizations as well.} 
\textit{Pan} moves more gently while maintaining a wider and stable view (\revision{see \autoref{fig:workspace}: \dottedlabel{Tracking in pan mode}}).
This panel provides four transitions as well as animation speed for shifting between global and detailed views (\roundedDClabel{DC3}, \roundedDClabel{DC4}): \textit{hardcut} switches immediately to the new view, \textit{zoom lens} animates a continuous push-in, \textit{dissolve} blends for a softer transition, and \textit{wipe} introduces a broadcast-style effect.

\noindent \panellabel{d3} \textit{Comparison Settings} control how multiple swimmers are selected and compared, including two layouts. 
\textit{Side-by-Side} separates selected swimmers into aligned strips for direct comparison.
\textit{Overlay} places selected swimmers on a blurred background, preserving continuity with the original footage.
This panel also provides transitions (\textit{hardcut} and \textit{smooth}) and animation duration for introducing comparison views (\roundedDClabel{DC3}, \roundedDClabel{DC4}). In addition, a \textit{History Data} panel displays real-time data dashboard(s) of selected live data metrics for the chosen swimmers, supporting inspection of gaps and trends.

\noindent \panellabel{d4} \textit{Layer Settings} control the design configuration of the selected data layer. It includes options for representation, placement, appearance, and layer-specific parameters. These settings determine how a layer is shown, where it is placed, and which specific options it uses, such as active lanes or reference targets. 


\vspace{2pt}
\noindent\textbf{Export:}
After completion, creators can export their designs as a standalone video (with optional data dashboard(s)) and/or a configuration \texttt{.json} file. Our export captures all final settings, producing an augmented race video ready for sharing or further editing.

\section{Evaluating Narrative Authoring in Motion:\\A User Study}
\label{sec:study}

\revision{Rather than evaluating \SCe\ as a complete video-editing tool in terms of usability, functionality, or production readiness, we evaluate it as a technology probe to investigate the benefits and challenges of authoring narrative visualizations in motion.}
%
We received ethics approval for our study from our institution’s ethics review board (XJTLU, \textnumero\,ER-LRR-11000180720240828231725). 
We pre-registered our study on OSF (\href{https://osf.io/zf7a9/}{\texttt{osf\discretionary{}{.}{.}io\discretionary{/}{}{/}zf7a9\discretionary{/}{}{/}}}).

\subsection{Procedure, Main Task, and Apparatus}
We prepared all the questionnaires (available in the supp.\ materials) used in our study on LimeSurvey~\cite{LimeSurvey}. 
Participants first completed a digital consent form; only those who consented proceeded to the pre-questionnaire, which collected demographics, content-creation experience, video-editing experience, visualization/infographic literacy, and familiarity with swimming races. Participants were then introduced to \SCe\ through a short video (available in supplemental materials), followed by hands-on training with a 2024 Paris Olympics men's \m{50} freestyle final video. During training, they explored \SCe\ and could ask questions freely.
Participants then completed the main task: authoring a narrative video with \SCe\ based on our recorded and combined footage of the 2024 Paris Olympic women's \m{100} butterfly final (\mins{1} \snd{2}). They first had 3 minutes to preview, fast-forward, slow down, or navigate within the footage, followed by \mins{5} to draft and explain an authoring plan on paper while keeping access to the footage. They then had 15 minutes to create their video using \SCe.
After the authoring task, participants completed a post-questionnaire. They first orally described how their final video differed from their initial plan. They then rated five statements on a 7-point Likert scale (1: strongly disagree; 7: strongly agree): whether they could find the needed functions, found \SCe\ easy to use, felt \SCe\ supported their creative process, felt \SCe\ helped turn race data into a coherent video story, and were satisfied with their created content. Participants briefly explained each rating. Finally, they orally answered open-ended questions about their authoring process, including when they chose \textit{Overview}, \textit{Tracking}, or \textit{Comparison}, how they applied \textit{Transition} effects, and how they selected, showed, switched, or hid data. They were asked for general comments (optional) before ending.

Depending on participants' availability, sessions were conducted either in person or remotely. In person, participants accessed \SCe\ through a local server on our laptop (a MacBook Pro 14-inch with an Apple M1 Pro chip and 16\,GB RAM); remotely, they used a public server on their own desktops or laptops. Based on our internal tests and pilot study, each session took approximately \mins{60}.

\vspace{-4pt}
\subsection{Participants}
We recruited participants through social media (\eg, WeChat, RedNote), our institutional mailing list, and the authors' networks, targeting individuals with experience in content creation (\ie, vloggers), video editing (\ie, animators), or graphic design (\ie, designers); basic knowledge of visualization and swimming was considered a plus. Eligible participants were at least 18 years old, proficient in English, and did not have motion sickness. Participation was voluntary, and participants could withdraw at any time, with withdrawn data excluded from analysis. Each completed participant received a 12.5~\euro\ gift card, above the local minimum legal hourly wage.

We totally had 9 participants: 5/9 \female\ and 4/9 \male, aged 18--54 \adjustbox{valign=c}{\includegraphics[height = 2ex]{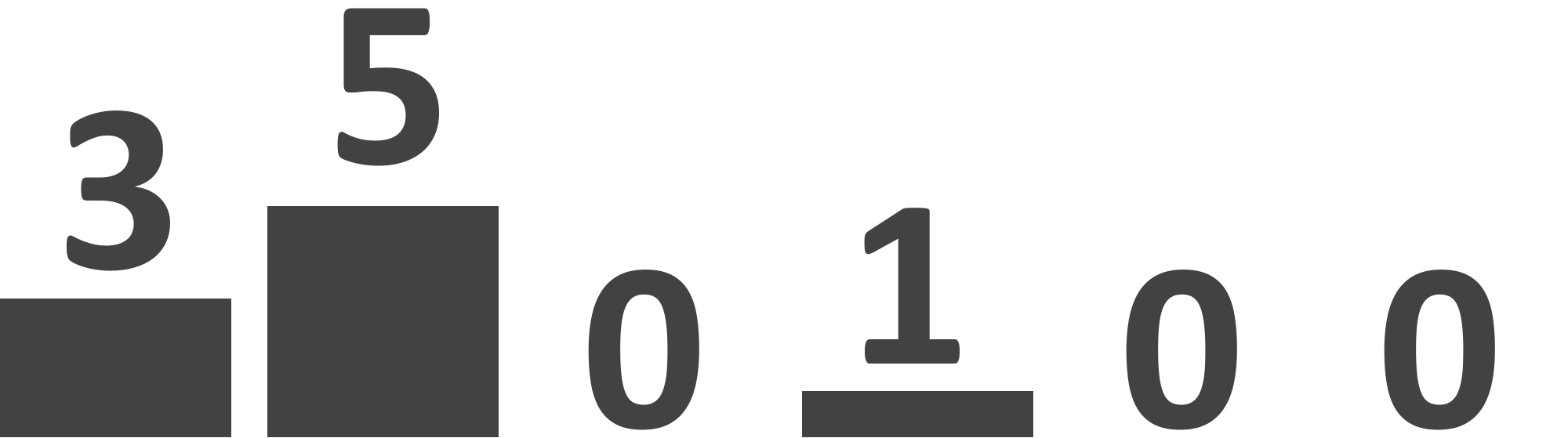}} \footnote{Age ranges: 18--24, 25--34, 35--44, 45--54, 55--64, and 65 or older.}. Participants had diverse backgrounds in visualization, content creation, sports, and professional media production, with several combining multiple roles. We report their main professions here, with details in the supplemental materials:
\percentagebar{0.44} 4/9 participants identified as content creators or vloggers, including a sports influencer with over 50,000 followers (\plabel{P5}), a vlogger (\plabel{P4}), a visualization design professor (\plabel{P9}), and an animation-design PhD student (\plabel{P6}). 
\percentagebar{0.67} 6/9 had professional or trained backgrounds in graphic/visual design, video editing, or visualization, including a professional video editor for national central television (\plabel{P8}), a graphic designer with 5+ years of experience (\plabel{P7}), a final-year visualization PhD student (\plabel{P2}), and an undergraduate visualization thesis student (\plabel{P1}). \plabel{P3} was a certified national master of sports in swimming.
\percentagebar{0.89} Almost all participants had content creation or video-editing experience: 6/9 had 5+ years (\plabel{P2}, \plabel{P4}, \plabel{P6}, \plabel{P7}, \plabel{P8}, \plabel{P9}), 2/9 had 2--5 years (\plabel{P1}, \plabel{P5}), and 1/9 had <1 year (\plabel{P3}). 
\percentagebar{0.67} More than half used video-editing tools at least sometimes: 1/9 daily (\plabel{P5}), 2/9 always (\plabel{P2}, \plabel{P6}), 3/9 sometimes (\plabel{P1}, \plabel{P4}, \plabel{P8}), 2/9 rarely (\plabel{P7}, \plabel{P9}), and 1/9 never (\plabel{P3}). 
\percentagebar{0.67} More than half had experience reading or creating visualizations/infographics: 4/9 regularly (\plabel{P1}, \plabel{P2}, \plabel{P6}, \plabel{P9}), 2/9 occasionally (\plabel{P4}, \plabel{P8}), and 3/9 little to none (\plabel{P3}, \plabel{P5}, \plabel{P7}).
\percentagebar{0.89} Almost all participants were familiar with swimming: 7/9 had watched some Olympic swimming races (\plabel{P1}, \plabel{P4}, \plabel{P5}, \plabel{P6}, \plabel{P7}, \plabel{P8}, \plabel{P9}), \plabel{P3} watched almost all races, and \plabel{P2} did not follow swimming races. 
Fewer participants (typically <3 per category; details in supp. materials) followed international, continental, national, or regional swimming competitions.

\vspace{-4pt}
\subsection{Data Collection and Analysis}
We collected participants' demographic and background responses, Likert-scale ratings, oral responses to open-ended questions, their draft plan written/sketched on paper, and the videos authored with \SCe. 
We audio-recorded and transcribed their open-ended responses for analysis.
For Likert-scale ratings, we counted the ratings for each item and descriptive statistics, including median, average, standard deviation, and rating distribution per Likert item.
For open-ended responses, one author initiated a codebook, coded the transcribed text for each question independently, then discussed ambiguous cases with another author. These two authors repeated this process until reaching a final agreement.
We conducted an observational analysis of all videos authored by our participants, from which we hope to understand the patterns underlying the narrative process of visualizations in motion.

\vspace{-4pt}
\subsection{Results}
We report results in two parts. The first includes quantitative data from Likert-scale responses and participants’ narrative choices and processes recorded in their \texttt{.json} files. The second presents qualitative findings derived from coding and observational analysis.  


\begin{figure}[tb]
  \centering
  \includegraphics[width=\columnwidth]{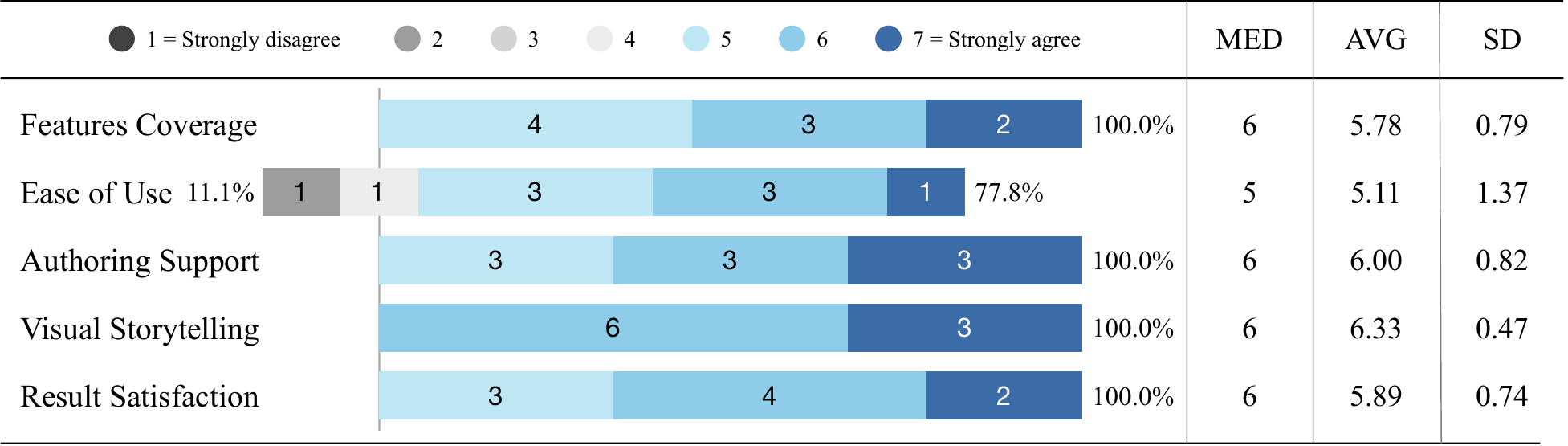}
  \caption{Distributions of participant ratings on \SCe’s feature coverage, ease of use, authoring support, visual storytelling support, and satisfaction with results on 7-point Likert scales, along with corresponding medians (MED), means (AVG), and standard deviations (SD).}
  \label{fig:quantitative}
\end{figure}

\vspace{2pt}
\noindent \textbf{Ratings on Likert Scales: }
Participants rated \SCe\ positively across all five 7-point Likert items (\autoref{fig:quantitative}). Features coverage, authoring support, visual storytelling, and results satisfaction were consistently rated above neutral, with high medians and averages and low variance. For ease of use, two participants rated it at or below neutral; reasons are discussed in the qualitative analysis. 

\begin{figure}[t]
  \centering
  \includegraphics[width=\columnwidth]{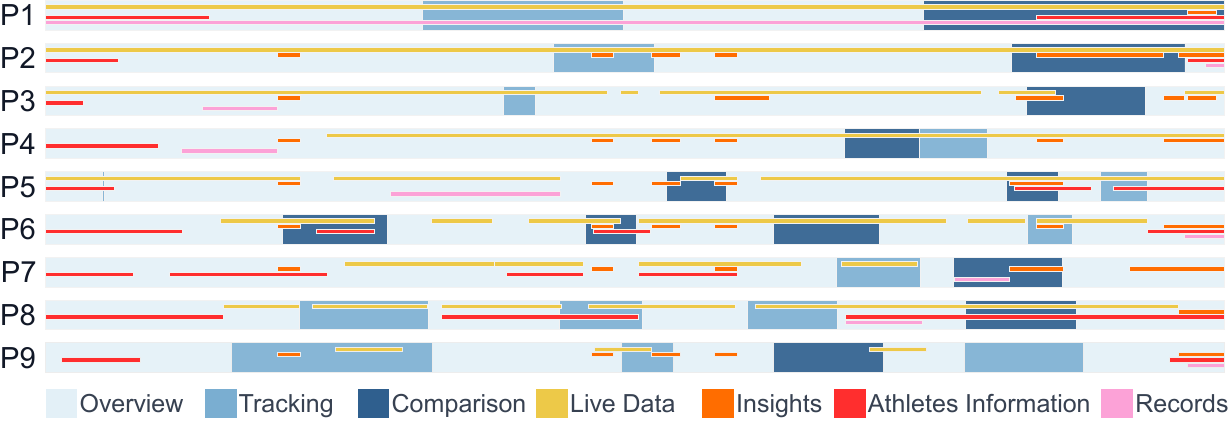}
  \caption{Authoring patterns of views and data layers over time.}
  \label{fig:test-pattern}
\end{figure}

\vspace{2pt}
\noindent \textbf{Observed Authoring Patterns:}
We present participants’ authoring patterns of views and data layers over time in \autoref{fig:test-pattern}. We report transition patterns and concrete statistics for views and data layers below:

\vspace{1pt}
\noindent \textit{Views:} On average, participants spent \snd{46.6} in \emph{Overview}, \snd{7.3} in \emph{Tracking}, and \snd{8.0} in \emph{Comparison}, confirming that the global view served as the main storytelling vehicle while the other two modes functioned as emphatic devices for key moments.

\vspace{1pt}
\noindent \textit{Transitions:} Since \emph{Zoom Lens} and \emph{Smooth} were default settings, all 9 participants used them. 2/9 participants used other transitions: \plabel{P5} used \emph{Wipe} once, \plabel{P8} used \emph{Hardcut} twice, and no participant used \emph{Dissolve}.

\vspace{1pt}
\noindent \textit{Data Layers:} \textit{Live data} 
accounted for \percentagebar{0.611} 61.1\% of the total authored layer duration, followed by \textit{athlete information} \percentagebar{0.221} 22.1\%, \textit{records} (9.2\%), and \textit{insights} (7.7\%).


\vspace{2pt}
\noindent \textbf{Rating Explanations': }
We report participants’ explanations for their ratings. Participants could express multiple opinions and comment on both the reasons for their ratings and any deductions. As such, the following results are based on coded open-ended responses and may not directly match Likert-scale counts; totals may also exceed 9. We present the findings in the order of the questions:

\vspace{1pt}
\noindent\textit{Narrative Plan Changes:} 
\percentagebar{0.67} 6/9 participants adjusted their initial plans when editing, while the remaining 3 mostly followed their original plans.
Changes included 
adding data layers (\texttimes\ 4; \eg, \plabel{P1} added \emph{Acceleration} near the race end to convey sprint tension), 
adjusting temporal arrangement (\texttimes\ 5; \eg, \plabel{P5} added more focus segments during the final chase among the top three swimmers),
and shifting their narrative focus (\texttimes\ 3; \eg,  \plabel{P7} expanded the introduction from the Chinese swimmer to the leader, while \plabel{P8} highlighted the American instead of the planned Australian swimmer). 
Additionally, \plabel{P8} wanted to include \emph{Personal Record} data, with his own designed representation.

\vspace{1pt}
\noindent \textit{Features Coverage:}
\percentagebar{1.00} All 9 participants said the available data and features were sufficient to produce coherent narrative videos. 
For example, \plabel{P4} found the data comprehensive and supported authoring content suitable for general audiences, and \plabel{P9} said our core functions met expectations for sports broadcasting.
While \percentagebar{0.55} 5/9 participants reported missing some professional editing features, including overlay layers, masking, and more flexible layer stacking (\plabel{P2}), keyboard shortcuts and timestamp tick marks (\plabel{P8}).
In addition, \plabel{P3} described the interface as crowded, and \plabel{P7} needed to test labels to understand their functions.
\revision{We discuss the limitations on functionalities in \autoref{sec:limitations}.}

\vspace{1pt}
\noindent \textit{Ease of Use:}
\percentagebar{0.56} 5/9 participants reported that our technology probe became reasonably usable after brief exploration. 
For example, \plabel{P1} found the interface complex at first but became easier over time, while \plabel{P2} and \plabel{P3} quickly understood and used the features.
In contrast, \percentagebar{0.22} 2/9 participants (\plabel{P4}, \plabel{P6}) reported a high initial learning burden due to the number of features. \percentagebar{0.33} 3/9 participants (\plabel{P7}--\plabel{P9}) noted mismatches with familiar video-editing interfaces, and preferred more direct timeline control.
\revision{We further discuss the usabilities in \autoref{sec:limitations}.}

\vspace{1pt}
\noindent \textit{Authoring Support:}
\percentagebar{0.56} 5/9 participants said that \SCe\ supported their authoring process by providing inspiration, particularly through \emph{Insights}.
\plabel{P2} described \emph{Insights} as effective in directing attention and establishing an authoring starting point. \plabel{P5} noted that some features actively triggered new narrative directions during editing.
\percentagebar{0.44} 4/9 participants said our features sufficiently support their authoring process: 
For example, \plabel{P1} said that \SCe\ supported her intended race explanation, and \plabel{P7} stated that it met most of her authoring needs. 
\percentagebar{0.22} 2/9 participants (\plabel{P4}, \plabel{P8}) noted that authoring support was contingent on overcoming an initial learning hurdle. As \plabel{P4} explained, \SCe\ became supportive after he understood the features. 

\vspace{1pt}
\noindent \textit{Visual Storytelling:}
\percentagebar{0.78} 7/9 participants reported that the available data were sufficient, and multiple views supported the visual storytelling of the race.
\plabel{P3} noted that the data clearly conveyed key aspects, \eg, speed, acceleration, turns, and gaps. \plabel{P5} found that the visualizations made race information easier to present to viewers. \plabel{P9} described the embedded visualizations, combined with view changes and lane comparisons, helped structure the race narrative.
\percentagebar{0.22} 2/9 participants said that pre-structured data reduced the authoring burden from the original video. \plabel{P2} noted that \SCe\ paired data with appropriate visualizations, allowing her to focus on layout and style rather than manual analysis. \plabel{P6} explained that, even without prior experience, \SCe\ helped her create content that looked clear and professional.

\vspace{1pt}
\noindent \textit{Result Satisfaction:}
All participants expressed overall satisfaction with their final videos.
Two participants explicitly compared their results with those achievable in professional video editing tools. 
\revision{\plabel{P5} was quite satisfied, noting that recreating the same visualizations in standard video editing tools would require substantially more time and effort. 
Similarly, \plabel{P7} thought that \SCe\ produced better results than attempting to achieve the same effects in professional video editing tools.}
\plabel{P4} was particularly satisfied, saying that the video clearly conveyed the swimmer's background, the first \m{50} performance, and final sprint in a way that general audiences could follow.
Two participants said they would like to further polish their videos. 

\vspace{2pt}
\noindent \textbf{Authoring Strategies:} 
Below, we report participants' answers to open-ended questions on their authoring strategies: 

\vspace{1pt}
\noindent \textit{Views:}
\percentagebar{1.00} All participants used \emph{Overview} to maintain global race context. For example, \plabel{P4} used it at the race start to help audiences understand the race as a whole, while \plabel{P8} found it essential for conveying overall standings.
\percentagebar{1.00} All participants used \emph{Tracking} to focus on specific moments or swimmers: \plabel{P1} used it near turns to closely follow leading swimmers, and \plabel{P5} used it when front swimmers were close to each other.
\percentagebar{1.00} All participants used \emph{Comparison} to clarify the race end: \plabel{P3} applied it near the race finish to compare leading swimmers and highlight medal competition.
\percentagebar{0.22} Two participants (\plabel{P1}, \plabel{P8}) mentioned shifting between \emph{Overview} and \emph{Tracking} to control narrative focus.

\vspace{1pt}
\noindent \textit{Transitions:}
\percentagebar{0.56} 5/9 participants explicitly described \emph{Zoom Lens} as effective for drawing attention to details at specific moments. 
For example, \plabel{P4} found it visually appealing without introducing a strong sense of discontinuity.
\percentagebar{0.44} 4/9 participants responded positively to \emph{Smooth}, describing it as supporting continuity; \plabel{P2} found it more comfortable and less abrupt than hardcuts.
\percentagebar{0.33} Three participants (\plabel{P6}, \plabel{P8}, \plabel{P9}) treated transitions as secondary to deciding what to show and when to shift. As \plabel{P9} explained, given the limited session time, his priority was shaping the narrative rather than exploring transition styles.

\vspace{1pt}
\noindent \textit{Data Layers:}
\percentagebar{0.78} 7/9 participants followed a similar event-driven strategy: they began with \emph{Athlete Information} (\eg, \emph{athlete name}, \emph{nationality}) to establish context, transitioned to \emph{Live Data} (\eg, \emph{current speed}, \emph{rank}, \emph{acceleration}) during the race, and returned to \emph{Athlete Information} and \emph{result} layers near the finish. 
For example, \plabel{P3} started with \emph{nationality} and \emph{athlete name}, showed \emph{current speed} in the first half, highlighted \emph{acceleration} before the turn, and referenced the gap to the \emph{world record} near the end. 
\plabel{P8} kept \emph{rank} visible for most of the race, saying the turn state was a key moment, and put the \emph{result} at the end. 
\percentagebar{0.22} Two participants (\plabel{P2}, \plabel{P6}) adopted a more data-driven authoring strategy. \plabel{P2} said she relied primarily on \emph{Insights}, then adjusted the layout and style of other layers to emphasize selected moments.

\vspace{2pt}
\noindent\textbf{General Comments:} We collected participants' general comments for both our technology probe and our study (optional questions):

\vspace{1pt}
\noindent \textit{To Our Technology Probe:}
\percentagebar{0.56} 5/9 participants described \SCe\ as already strong. For example, \plabel{P9} found it well designed for the specific use case of post-editing and replaying of race videos in sports broadcasting. \plabel{P2} said that it substantially reduced the effort required for race analysis.
\percentagebar{0.56} 5/9 participants also provided suggestions, mainly on usability and layout. \plabel{P4} suggested a more intuitive feature organization for broader users. \plabel{P8} noted that although \SCe\ aligns well with swimming data, usability could be further improved. \plabel{P2} additionally requested more visual representation options.

\vspace{1pt}
\noindent \textit{To Our Study:}
\percentagebar{0.33} 3/9 participants (\plabel{P2}, \plabel{P8}, \plabel{P9}) said our study was overall great. \percentagebar{0.22} 2/9 participants (\plabel{P1}, \plabel{P8}) reported no concerns.  \percentagebar{0.33} 3/9 participants (\plabel{P4}, \plabel{P7}, \plabel{P9}) said practically, they would not make a plan before editing. Specifically, \plabel{P4} preferred planning while editing.

\textbf{Overall,} 
\revision{participants rated \SCe\ positively with some reported learning cost and missing features, which we further discuss in \autoref{sec:limitations}.}
Their authoring patterns were consistent: \emph{Overview} served as the main scaffold, complemented by \emph{Tracking} and \emph{Comparison}, while data layers followed an event-driven progression from metadata to live data and outcomes.
\emph{Insights} and structured data supported authoring by providing entry points, inspiration, and reduced analysis effort.

\section{Discussions and Limitations}
\label{sec:limitations}

\revision{
\noindent\textbf{Robustness in Functionality, Usability, and Narrative Expressiveness:}
Although participants rated \SCe\ positively overall, our results reveal a tension between \textit{ease of use} and \textit{feature coverage}. While \SCe\ is a technology probe, this tension is important to discuss as it points to a broader design challenge for complete tools.
%
Visualizations embedded in sports videos are not decorations; they encode dynamic data, often update rapidly, and need to remain perceptually and semantically connected to their data referents. 
Authoring visualizations in motion thus requires more than conventional video editing, including data-, design-, and motion-oriented controls \cite{Yao:2024:SwimmingVideos}. 
The embedding locations of visualizations in motion also need to be continuously adjusted; otherwise, visualizations may detach from the data referent or worse, convey misleading information.
This creates an inherent trade-off: richer controls may support more precise and flexible storytelling, but would also increase learning cost and interaction burden. We observed this in layer control: Prior tools like \textit{SwimFlow}~\cite{Yao:2024:SwimmingVideos} supported flexible layer stacking, but participants found it hard to manage and preferred direct timeline control. We simplified layer management in \SCe, yet experienced participants requested finer-grained control.
A similar tension appears in visual representation: although we provided dedicated swimming encodings~\cite{Yao:2024:SwimmingVideos}, some participants wanted to design their own.
However, designing readable visualizations in motion that evolve with moving referents is difficult even for experienced designers, and becomes harder for non-experts.
Concrete guidelines are thus required to deal with such trade-offs, for example, which features should be prioritized, what level of manipulation granularity is appropriate, and which interaction techniques best support each authoring operation. These questions remain open and are central to balancing functionality, usability, and narrative expressiveness when supporting visualizations in motion.
}

\revision{
\noindent\textbf{Shared Narrative Structures, Divergent Authoring Needs:}
Participants with diverse backgrounds independently developed a similar narrative structure: \emph{Overview} served as the backbone, while \emph{Tracking} and \emph{Comparison} were used for key moments. Data layers followed a similar event-driven progression, moving from athlete metadata to live race states, and finally to results. This consistency suggests that narrative authoring for visualizations in motion may depend more on the event's temporal logic than on participants' backgrounds.
Yet, participants' backgrounds shaped what they expected from \SCe. Participants with video-editing experience treated it as an editing tool and expected professional workflow features, while those in visualization/design focused more on representation and narrative choices, and audience guidance. Notably, \plabel{P3}, the professional swimmer with no editing/visualization background, strongly preferred text for its similarity to practical broadcasts. Thus, participants' backgrounds determined how they treated and used our probe and the features they cared about.
}

\revision{
\noindent\textbf{Open Challenges for Authoring Visualization in Motion:}
We work on swimming for its rich data and motion, near-linear trajectories that support tracking~\cite{Yao:2022:VIM}, and the availability of videos~\cite{Yao:2024:SwimmingVideos}. 
We started from a bird's-eye view, using homographic transformations to simulate alternative viewpoints and panning to simulate camera movement, allowing us to first study authoring visualizations in motion within a relatively stable spatial structure.
Our setup reveals what remains difficult. Real broadcast scenarios involve both moving athletes and moving cameras, where the solution is not simply to apply multi-homography transformations to visualizations in motion, as they need to be readable, appropriately embedded, and properly associated with their referents.
For example, when an athlete temporarily leaves the camera view, should the corresponding visualization disappear, remain as contextual information, or move to an extended display space, \ie, cross-reality? When athletes overlap in some perspectives, for example, when the camera points from the long side of the swimming pool, how should embedded visualizations avoid occlusion while preserving correct referents?
These questions point to broader challenges in dynamic layout, spatial embedding locations, and narrative continuity for visualizations in motion.
To explore the extendability of our work, we additionally authored a real Olympic running broadcast video with a moving camera, using our data preparation pipeline presented in this paper and \SCe\ features. We do not include the video here due to copyright restrictions, but provide it in our OSF repository (see supp.\ material).
}

\section{Conclusion}
We investigated how to author narrative visualizations in motion in swimming \revision{videos}. We proposed an automatic multimodal data preparation pipeline, conducted an observational analysis, implemented a technology probe to support the authoring process, and evaluated it through an empirical study. We discussed and proposed open research opportunities for \revision{authoring} visualizations in motion.

\section*{Supplemental Material Pointers}
\label{sec:supplemental_materials}

The pre-registration for our empirical study can be found at \href{https://osf.io/zf7a9/}{\texttt{osf\discretionary{}{.}{.}io\discretionary{/}{}{/}zf7a9\discretionary{/}{}{/}}}.
Respectively, we also share our
(a) source code for our data preparation pipeline presented in \autoref{sec:pipeline}, \revision{including our implemented LLM skills with prompts, schemas, and validation scripts}; 
(b) observational analysis materials, including original data and counts to support \autoref{sec:frombroadcasts};
(c) source code for our technology probe \SCe\ presented in \autoref{sec:swimcomposer};
(d) empirical evaluation results presented in \autoref{sec:study}, including our study draft, introductory video used in the study, participants' final authored videos, participants' original answers to questionnaires and our analysis scripts, participants' original answers to open-ended questions and our codebook as well as scripts;
(e) \revision{a demo authored from a real Olympic running race broadcast video with a moving camera by using our data preparation pipeline and authoring with narrative features in \SCe;}
and (f) other materials such as ethics approval at \href{https://osf.io/bq47n/}{\texttt{osf\discretionary{}{.}{.}io\discretionary{/}{}{/}bq47n\discretionary{/}{}{/}}}.

\SCe\ can be accessed at \href{http://43.163.231.237/}{\texttt{43\discretionary{}{.}{.}163\discretionary{}{.}{.}231\discretionary{}{.}{.}237\discretionary{/}{}{/}}}.

\section*{Images/graphs/plots/tables/data license/copyright}
\label{sec:figure_credits}

As authors, we state that all of our own figures, graphs, plots, and data tables in this article are and remain under our own personal copyright, with the permission to be used here. We also make them available under the \href{https://creativecommons.org/licenses/by/4.0/}{Creative Commons At\-tri\-bu\-tion 4.0 International (\ccLogo\,\ccAttribution\ \mbox{CC BY 4.0})} license and share them at \href{https://osf.io/bq47n/}{\texttt{osf\discretionary{}{.}{.}io\discretionary{/}{}{/}bq47n\discretionary{/}{}{/}}}.

\section*{Acknowledgements}
This work is partially supported by XJTLU RDF, grant \textnumero\ RDF-24-01-062; the Open Project Program of the State Key Laboratory of CAD\&CG (Grant \textnumero\ A2514), Zhejiang University; ANR, the French National Research Agency, with the SPORTSVIZ project, grant \textnumero\ ANR-24-CE33-4945.
We used a large language model (ChatGPT) to edit, enhance grammar, and improve clarity during the writing process.

\bibliographystyle{abbrv-doi-hyperref}

\bibliography{Junhao}

\end{document}